\documentclass[final,1p,authoryear]{elsarticle}

\usepackage{amssymb,amsmath,accents,enumitem}

\usepackage{float}

\usepackage{color}

\usepackage{array}

\usepackage[colorlinks,urlcolor=blue]{hyperref}

\newcommand{\vett}[1]{\mathbf{#1}}

\newcommand {\tr} {\mbox{\rm tr\,}}

\newcommand {\dd}{\mathrm{d}}
{\left\lbrace\begin{array}{@{}l@{}}}%
{\end{array}\right.}

\begin{document}

\begin{frontmatter}



\title{A geometrically exact model for thin magneto-elastic shells}


\author[label1]{Matteo Pezzulla\footnote{Current affiliation: Slender Structures Lab,
	Department of Mechanical and Production Engineering,
	{\AA}rhus University,
	Inge Lehmanns Gade 10, 8000 {\AA}rhus C, Denmark}}
\ead{matt@mpe.au.dk}

\author[label1]{Dong Yan}
\ead{dong.yan@epfl.ch}

\author[label1]{Pedro M. Reis\corref{cor1}}
\ead{pedro.reis@epfl.ch}

\cortext[cor1]{Corresponding author}
    
    \address[label1]{Flexible Structures Laboratory,
	Institute of Mechanical Engineering,
	\'{E}cole Polytechnique F\'{e}d\'{e}rale de Lausanne,
	1015 Lausanne, Switzerland
    }

\address{}

\begin{abstract}
We develop a reduced model for hard-magnetic, thin, linear-elastic shells that can be actuated through an external magnetic field, with geometrically exact strain measures. Assuming a reduced kinematics based on the Kirchhoff-Love assumption, we derive a reduced two-dimensional magneto-elastic energy that can be minimized through numerical analysis. In parallel, we simplify the reduced energy by expanding it up to the second order in the displacement field and provide a physical interpretation. Our theoretical analysis allows us to identify and interpret the two primary mechanisms dictating the magneto-elastic response: a combination of equivalent magnetic pressure and forces at the first order, and distributed magnetic torques at the second order. We contrast our reduced framework against a three-dimensional nonlinear model by investigating three test cases involving the indentation and the pressure buckling of shells under magnetic loading. We find excellent agreement between the two approaches, thereby verifying our reduced model for shells undergoing nonlinear and non-axisymmetric deformations. We believe that our model for magneto-elastic shells will serve as a valuable tool for the rational design of magnetic structures, enriching the set of reduced magnetic models. 

\end{abstract}


%
%
%


\end{frontmatter}



\section{Introduction}
\label{sec:introduction}

Investigating the effect of non-mechanical stimuli on structures has long been the subject of many research efforts, from the bending of bi-metallic thermostats \citep{Timoshenko1925} to the swelling-induced morphing of thin structures \citep{Kim2012}. While these stimuli can induce extreme deformations, the time scales of the actuation processes can be relatively large, in the order of some seconds or minutes, a feature that might not be ideal in tasks where fast actuation is required. By contrast, the coupling between magnetism and elasticity represents a valuable avenue towards fast and reversible actuation of soft structures \citep{Lum_PNAS2016,Kim_Nature2018,Zhao2019,Wang_JMPS2020}. A promising class of materials for such actuations is that of magneto-rheological elastomers (MREs), which are composites of magnetized (metallic) particles and a soft elastic polymeric matrix that respond to an external applied magnetic field. Applications of MREs range from minimal invasive procedures \citep{Pancaldi2020} to soft robotic actuators \citep{Hu2018,Gu_NatCommun2020,alapan_reprogrammable_2020} and bio-medical devices \citep{Kim_SciRobot2019}.

Depending on the extent to which MREs can keep their residual magnetization and how they respond to an external magnetic field, these materials can be classified into different categories, such as superparamagnetic and soft-ferromagnetic materials. For a more detailed discussion on these different categories, we refer to \citep{Bertotti_magnetism_1998,Sano2020prep}. Past pioneering studies have primarily focused on the deformation of structures made of superparamagnetic~\citep{moon_magnetoelastic_1968,cebers_dynamics_2003,cebers_bending_2004,cebers_magnetic_2007,dreyfus_microscopic_2005,roper_dynamics_2006,gerbal_refined_2015} or soft-ferromagnetic~\citep{rigbi_response_1983,ginder_magnetorheological_1999,dorfmann_magnetoelastic_2003,danas_experiments_2012,Loukaides_IntJSmartNanoMater2014,Seffen_SmartMaterStruct2016,Psarra_JMPS2019} materials under an external magnetic field. More recent efforts have turned to hard-magnetic soft materials~\citep{Lum_PNAS2016,Kim_Nature2018,Zhao2019,Wang_JMPS2020,Yan_NatureMaterials_2020}. These materials are MREs made of hard-ferromagnetic particles embedded in a soft elastomeric matrix. They retain a permanent magnetization, have high coercivity (\emph{i.e.} the necessary field strength to erase the magnetization) upon saturation, while being mechanically compliant due to the soft elastomeric matrix and their usual slender geometry utilized in applications. Therefore, hard MREs present several advantages for tasks where a fast and reversible actuation is required, as for example in soft robotics, minimal invasive procedures, and bio-medical devices.

During the past decade, there have been several efforts on the modeling of these materials, motivated by the numerous possible applications of hard MREs~\citep{danas_experiments_2012}. As an example, a nonlinear three-dimensional (3D) theory for hard MREs was proposed by \cite{Zhao2019}, where the Helmholtz free energy of the system consists of elastic (neo-Hookean) and magneto-elastic contributions. The basic assumption, following the physical observations in \citep{Bertotti_magnetism_1998}, is that the magnetic flux density is linear with the applied magnetic field. The model was then tested against experimental results, by implementing the theoretical model in a finite element scheme in Abaqus, resulting in a quantitative agreement between the two.
 
Based on the 3D model presented in \citep{Zhao2019}, reduced theories for hard magnetic linear and nonlinear beams  \citep{Wang_JMPS2020,Yan2020prep}, and rods \citep{Sano2020prep} have been derived. In these studies, a dimensional reduction procedure was performed on the 3D magneto-elastic energy, assuming a reduced kinematics for the beams and the rods based on the Kirchhoff-Love assumptions. Moreover, we have recently presented a study on magneto-active axisymmetric shells made of hard MREs in \citep{Yan_NatureMaterials_2020}, where the coupling between mechanics and magnetism was leveraged to change the stability onsets of shells undergoing pressure buckling \citep{hutchinson_imperfections_2018,lee_geometric_2016,lee_evolution_2019}. In this study, experiments were contrasted with a magnetic shell model for axisymmetric deformation and geometrically exact strain measures \citep{Yan_NatureMaterials_2020}. The results show that the magnetic field can be used to tune the critical buckling pressure of spherical shells, which are highly sensitive to imperfections \citep{hutchinson_effect_1971,hutchinson_john_w._buckling_2016,hutchinson_imperfections_2018}. Theoretical models for non-axisymmetric deformations of magneto-active shells have received less attention and have been derived for shallow shells only \citep{Seffen_SmartMaterStruct2016,Loukaides_IntJSmartNanoMater2014}, with an understanding of the coupling between magnetism and mechanics in the general case that remains unclear. In our view, deriving reduced-order structural models should have the twofold goal of providing alternatives to 3D models that can be solved numerically with a reduced computational cost, and enabling a better understanding of the mechanics of the problem given the lower complexity of the reduced models.

In this paper, we derive a theory for thin, elastic, magnetic shells with geometrically exact strain measures \citep{babcock_shell_1983,Koiter1969,Niordson1985}, thereby generalizing the model that we presented in \citep{Yan_NatureMaterials_2020} that was only limited to axisymmetric shells undergoing axisymmetric deformations. A model with even simpler kinematics, such as with moderate rotations \citep{Sanders1963,Donnell1977}, could also be derived to further reduce the complexity of the model, but this derivation is beyond the scope of the paper as we want to focus on the more general formulation with geometrical exact strain measures. In particular, we perform a dimensional reduction of the 3D magneto-elastic energy contribution presented in \citep{Zhao2019}, by assuming a reduced kinematics according to the Kirchhoff-Love assumptions \citep{Niordson1985}. The result is a reduced bidimensional (2D) energy that can be minimized via a finite element scheme implemented in COMSOL Multiphysics, a commercial software. Moreover, given that the reduced magnetic energy is highly nonlinear and not amenable to physical interpretation, we expand it up to the second order in the displacement field and untangle two different contributions with a clear physical meaning. At the first order in the displacement field, the magneto-elastic coupling can be interpreted as a combination of equivalent magnetic pressure and in-plane forces. At the second order, the magneto-elastic coupling is instead represented by distributed torques. To verify our reduced model, we test it against the existing 3D model derived by \citep{Zhao2019}, implemented in Abaqus, finding excellent agreement in a set of three different test cases that we investigate in detail. Specifically, we test the model in the cases of (i) non-axisymmetric indentation under a magnetic field; (ii) pressure buckling under asymmetric magnetic loading where the residual magnetization vector and the external magnetic field are in the same plane; and (=iii) pressure buckling under asymmetric magnetic loading, where the residual magnetization vector and the external magnetic field are not in the same plane.

Our paper is organized as follows. In Sec. \ref{sec:preliminaries}, we set the notation, describe the geometry of the shell, and recall the reduced elastic energy of its mid-surface. We perform the dimensional reduction on the magnetic energy of the shell in Sec. \ref{sec:dimensionalreduction}. Then, in Sec. \ref{sec:interpretation}, we provide a mechanical interpretation to the reduced magnetic energy. The numerical implementation of our model is summarized in Sec. \ref{sec:numerical}, whereas the validation with the three different problems is presented in Sec. \ref{sec:validation}. In Sec. \ref{sec:conclusion}, we provide final discussions where we also summarize our findings.
 
\section{Preliminaries}
\label{sec:preliminaries}

For the sake of completeness and convenience to the reader, we start by recalling some standard concepts of shell theory~\citep{Niordson1985} and differential geometry~\citep{Oneill1997,docarmo2016} that we will need for the analysis presented in this paper, since this material is often difficult to find in a synthesized way in the vast and fragmented literature of shell mechanics. We will also recall the reduced energy for linearly elastic thin shells, with geometrically exact strain measures.

For the rest of the Section, we will follow mainly two references: the book on shell theory by~\cite{Niordson1985}, and the book on elementary differential geometry by~\cite{Oneill1997}. When dealing with entities from 3D continuum mechanics, we will follow the notation of~\cite{Gurtin2010}.

\subsection{Geometry of the shell}

\begin{figure}[h]
    \centering
    \includegraphics[scale=1.1]{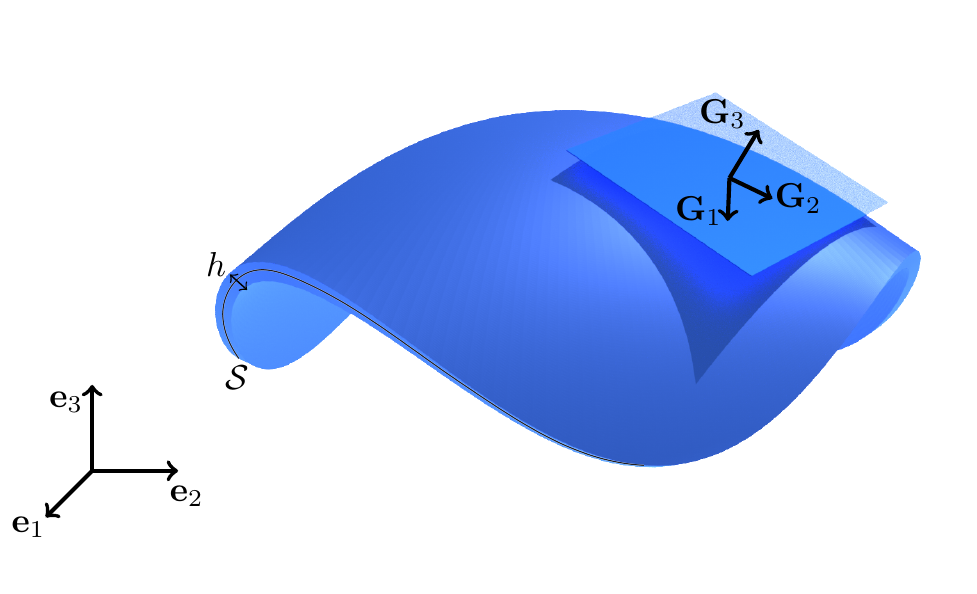}
    \caption{Schematic representation of a non-axisymmetric shell, in its reference configuration. The covariant base vectors~$\mathbf{G}_i$ are drawn at one point of the top surface, with the tangent plane. The thickness~$h$ and the mid-surface~$\mathcal{S}$ are also shown.}
    \label{fig:shell}
\end{figure}

A shell is a 3D body, $\mathcal{B}$, embedded in an Euclidean space, $\mathcal{E}$, with Cartesian basis~$(\mathbf{e}_1,\mathbf{e}_2,\mathbf{e}_3)$. One of the characteristic dimensions of a shell (the thickness) is much smaller than the other two (along the surface), as depicted in the schematic diagram of Fig.~\ref{fig:shell}. Shells are typically described by their mid-surface~$\mathcal{S}\subset\mathcal{E}$, curved in its natural (stress-free) state. We denote the parametrization of the mid-surface as~$\accentset{\circ}{\mathbf{r}}(\eta^1,\eta^2)\colon\mathcal{D}\rightarrow\mathcal{E}$. Here, $(\eta^1,\eta^2)$ are curvilinear coordinates, $\mathcal{D}$ is the domain of parametrization, and the accent $\accentset{\circ}{()}$ denotes quantities in the \textit{undeformed configuration}. As a 3D body, the shell is a stack of surfaces~\citep{Oneill1997}, viewed as the Cartesian product~$\mathcal{B}=\mathcal{S}\times[-h/2,h/2]$, where~$h(\eta^1,\eta^2)$ is the thickness of the shell, which is, in general, a function of the curvilinear coordinates.

Once the parametrization of the mid-surface is given, the covariant base vectors associated to the surface can be computed as~$\accentset{\circ}{\mathbf{a}}_{\alpha}=\accentset{\circ}{\mathbf{r}},_{\alpha}$, where~$(),_{\alpha}$ denotes partial differentiation with respect to~$\eta^{\alpha}$, and Greek indices run from~$1$ to~$2$ \citep{Oneill1997}. The vectors~$\accentset{\circ}{\mathbf{a}}_{\alpha}$ give rise to a covariant basis for the mid-surface~$\mathcal{S}$, whose covariant metric coefficients can be determined as~$\accentset{\circ}{a}_{\alpha\beta}=\accentset{\circ}{\mathbf{a}}_{\alpha}\cdot\accentset{\circ}{\mathbf{a}}_{\beta}$, also referred to as first fundamental form, where~$(\cdot)$ denotes the inner product in the Euclidean space~$\mathcal{E}$. The contravariant metric coefficients can be determined as the components of the inverse matrix~$\accentset{\circ}{a}^{\alpha\beta}=(\accentset{\circ}{a}_{\alpha\beta})^{-1}$, so that the contravariant base vectors can be computed as~$\accentset{\circ}{\mathbf{a}}^{\alpha}=\accentset{\circ}{a}^{\alpha\beta}\accentset{\circ}{\mathbf{a}}_{\beta}$ \citep{Oneill1997}. Finally, the normal unit vector to the mid-surface is computed as~$\accentset{\circ}{\mathbf{n}}=(\accentset{\circ}{\mathbf{a}}_1\times\accentset{\circ}{\mathbf{a}}_2)/|\accentset{\circ}{\mathbf{a}}_1\times\accentset{\circ}{\mathbf{a}}_2|$, ensuring its perpendicularity to any tangent plane along the mid-surface. Furthermore, the covariant components of the curvature tensor are defined as~$\accentset{\circ}{b}_{\alpha\beta}=\accentset{\circ}{\mathbf{n}}\cdot\accentset{\circ}{\mathbf{r}},_{\alpha\beta}$ (also referred to as second fundamental form).

With the definition of the normal vector $\accentset{\circ}{\mathbf{n}}$, the parametrization of the shell body~$\mathcal{B}$ can be written as~$\mathbf{X}=\accentset{\circ}{\mathbf{r}}+\eta^3\accentset{\circ}{\mathbf{n}}$, where~$\eta^3$ is the coordinate running along the normal vector. Then, with Latin indices run from~$1$ to~$3$, the covariant base vectors associated with~$\mathbf{X}$ are~$\mathbf{G}_i=\mathbf{X},_i$. By applying the definition of the covariant base vectors~$\mathbf{G}_i=\mathbf{X},_i$ to the parametrization~$\mathbf{X}$, one can derive
\begin{equation}
\begin{aligned}
    &\mathbf{G}_{\alpha}=\accentset{\circ}{\mathbf{a}}_{\alpha}+\eta^3\accentset{\circ}{\mathbf{n}},_{\alpha}\,,\\
    &\mathbf{G}_3=\accentset{\circ}{\mathbf{n}}\,.\\
    \end{aligned}
    \label{eq:undefcov}
\end{equation}
Similarly to what was derived for the mid-surface (\emph{i.e.} the first and second fundamental forms $\accentset{\circ}{a}_{\alpha\beta}$ and $\accentset{\circ}{b}_{\alpha\beta}$), the 3D covariant metric coefficients can be computed as~$G_{ij}=\mathbf{G}_i\cdot\mathbf{G}_j$, and the contravariant metric coefficients as~$G^{ij}=(G_{ij})^{-1}$. Then, the contravariant base vectors can be computed as~$\mathbf{G}^i=G^{ij}\mathbf{G}_j$. 

Finally, when dealing with a stack of surfaces, it is natural to ask how the area measure varies along the thickness coordinate~$\eta^3$. Not only is this a remarkable result in differential geometry, but is also relevant to any recipe for dimensional reduction involving shells. Defining~$\accentset{\circ}{G}=\sqrt{\det(G_{ij})}$ and~$\accentset{\circ}{a}=\sqrt{\det(\accentset{\circ}{a}_{\alpha\beta})}$, it can be shown that~\citep{Oneill1997}
\begin{equation}
    \accentset{\circ}{G}=(1-2\eta^3\accentset{\circ}{H}+(\eta^3)^2\accentset{\circ}{K})\accentset{\circ}{a}\,,
    \label{eq:jacobian}
\end{equation}
where~$\accentset{\circ}{H}$ and~$\accentset{\circ}{K}$ are the mean and Gaussian curvatures of the undeformed mid-surface, respectively. This geometric result explains how local areas vary within the stack of surfaces, when moving along the normal. For thin shells, that is at first order in $\eta^3$, changes in local areas are dictated by the mean curvature only, which is why minimal surfaces have zero mean curvature \citep{Deserno2004}.

When the shell deforms in response to mechanical loads (\textit{e.g.}, distributed forces and pressures) or non-mechanical loads (\textit{e.g.}, swelling or magnetic fields), its material points are displaced, forming a new configuration, termed \textit{deformed} or \textit{actual configuration}. The positions of the material points in this new configuration are denoted as~$\mathbf{x}$, as opposed to~$\mathbf{X}$ for the undeformed configuration. For thin shells, a reduced kinematics is usually assumed following the Kirchhoff-Love assumptions, according to which fibers normal to the mid-surface do not stretch nor shrink, and remain normal to the mid-surface, upon deformation. Assuming such a reduced kinematics means that the 3D deformed shape of the shell can be fully described by the sole shape of its mid-surface.

All the quantities that have been computed in the undeformed state such as the base vectors $\mathbf{G}_i$ and the fundamental forms $\accentset{\circ}{a}_{\alpha\beta}$ and $\accentset{\circ}{b}_{\alpha\beta}$, defined by the parametrization~$\mathbf{X}$, can be now expressed in the deformed state, $\mathbf{x}$. The parametrization of the deformed mid-surface is denoted as~$\mathbf{r}=\accentset{\circ}{\mathbf{r}}+\mathbf{u}$, where~$\mathbf{u}$ is the displacement of each material particle of the mid-surface, from the undeformed to the deformed configurations. The displacement field is usually expressed in the covariant basis as~$\mathbf{u}=u^\alpha\accentset{\circ}{\mathbf{a}}_{\alpha}+w\accentset{\circ}{\mathbf{n}}$, where Einstein's summation convention is implied and $\check{\mathbf{u}}=u^\alpha\accentset{\circ}{\mathbf{a}}_{\alpha}$ is the in-plane displacement vector field. The covariant base vectors of the deformed mid-surface are~$\mathbf{a}_{\alpha}=\mathbf{r},_{\alpha}$, whereas those associated with~$\mathbf{x}$ are~$\mathbf{g}_i=\mathbf{x},_i$. The normal vector is~$\mathbf{n}=(\mathbf{a}_1\times\mathbf{a}_2)/|\mathbf{a}_1\times\mathbf{a}_2|$. Finally, the first and second fundamental forms associated with the deformed mid-surface are, respectively, $a_{\alpha\beta}=\mathbf{a}_{\alpha}\cdot\mathbf{a}_{\beta}$ and~$b_{\alpha\beta}=\mathbf{n}\cdot\mathbf{r},_{\alpha\beta}$. Similarly to what derived for the undeformed shell in Eq.~\eqref{eq:undefcov}, the base vectors~$\mathbf{g}_i$ can be expressed as
\begin{equation}
\begin{aligned}
    &\mathbf{g}_{\alpha}=\mathbf{a}_{\alpha}+\eta^3\mathbf{n},_{\alpha}\,,\\
    &\mathbf{g}_3=\mathbf{n}\,.\\
    \end{aligned}
    \label{eq:defcov}
\end{equation}

With classic formalism describing the geometry of the shell presented above, we can now move on to specify the deformation gradient and its expression for thin shells. 

\subsection{The deformation gradient}

As it will be made clear in Sec.~\ref{sec:dimensionalreduction}, the deformation gradient~$\mathbf{F}$ has to be expressed in terms of the quantities describing the shell kinematics to be able to perform the dimensional reduction. 

The deformation gradient is a two-point tensor that maps the undeformed tangent space to the deformed tangent space, defined as~\citep{Gurtin2010}
\begin{equation}
    \mathbf{F}=\mathbf{g}_i\otimes\mathbf{G}^i=\mathbf{g}_{\alpha}\otimes\mathbf{G}^{\alpha}+\mathbf{n}\otimes\accentset{\circ}{\mathbf{n}}\,.
    \label{eq:F3D}
\end{equation}
Before we can express $\mathbf{F}$ only in terms of surface quantities, we first need to quantify how the contravariant components of the 3D metric~$G^{ij}$ vary as a function of the thickness coordinate~$\eta^3$. Expressing ~$G^{ij}$ as a function of~$\eta^3$ is a common procedure for plates, where the deformation gradient is a linear function of the thickness coordinate~\citep{Pietraszkiewicz1980}, whereas it is more intricate for shells because of the existing rest curvature. 

Before we can compute the contravariant metric coefficients of the undeformed shell~$G^{\alpha\beta}$, we first compute the covariant components as~\citep{Oneill1997}
\begin{equation}
G_{\alpha\beta}=\mathbf{G}_{\alpha}\cdot\mathbf{G}_{\beta}=\accentset{\circ}{a}_{\alpha\beta}-2\eta^3\accentset{\circ}{b}_{\alpha\beta}+(\eta^3)^2\accentset{\circ}{c}_{\alpha\beta}\,,
\label{eq:galphabeta}
\end{equation}
where~$\accentset{\circ}{c}_{\alpha\beta}=\accentset{\circ}{\mathbf{n}},_{\alpha}\cdot\accentset{\circ}{\mathbf{n}},_{\beta}$ is the third fundamental form of the undeformed mid-surface. Recalling that, by definition, $\mathbf{G}_3\cdot\mathbf{G}_{\alpha}=0$ and that~$\mathbf{G}_3\cdot\mathbf{G}_3=1$, the covariant 3D metric coefficients of the undeformed shell are
\begin{equation}
G_{ij}=
\begin{pmatrix}
    G_{\alpha\beta} & \begin{matrix} 0 \\ 0 \end{matrix} \\
    \begin{matrix} 0 & 0 \end{matrix} & \!\!1
\end{pmatrix}\,.
\end{equation}
The structure of the metric conveys that the normal direction, assumed to be a coordinate line, is indeed normal to the mid-surface. Since~$G_{\alpha3}=0$, we have that~$G^{\alpha3}=0$ and~$G^{\alpha\beta}=(G_{\alpha\beta})^{-1}$. To perform this inversion and simplify the algebra, we employ matrix notation, where the covariant components of the first, second, and third fundamental forms are denoted as~$\mathbb{A}$, $\mathbb{B}$ and~$\mathbb{C}$ so that the covariant 3D metric may be expressed as
\begin{equation}
\mathbb{G}=\mathbb{A}-2\eta^3\mathbb{B}+(\eta^3)^2\mathbb{C}=\mathbb{A}(\mathbb{I}-2\eta^3\mathbb{A}^{-1}\mathbb{B}+(\eta^3)^2\mathbb{A}^{-1}\mathbb{C})\,,
\end{equation}
where~$\mathbb{I}$ is the identity matrix. We compute the inverse matrix of~$\mathbb{G}$ as
\begin{equation}
\mathbb{G}^{-1}=(\mathbb{I}-2\eta^3\mathbb{A}^{-1}\mathbb{B}+(\eta^3)^2\mathbb{A}^{-1}\mathbb{C})^{-1}\mathbb{A}^{-1}\,,
\end{equation}
and expand it up to second order in~$\eta^3$ as
\begin{equation}
\mathbb{G}^{-1}=\mathbb{A}^{-1}+2\eta^3\mathbb{A}^{-1}\mathbb{B}\mathbb{A}^{-1}-(\eta^3)^2\mathbb{A}^{-1}\mathbb{C}\mathbb{A}^{-1}+4(\eta^3)^2(\mathbb{A}^{-1}\mathbb{B})^2\mathbb{A}^{-1}+O((\eta^3)^3)\,.
\label{eq:Gcontramatrix}
\end{equation}
Returning to index notation, we notice that a pre-multiplication of a tensor by~$\mathbb{A}^{-1}$ corresponds to raising the first index of that tensor, whereas a post-multiplication by~$\mathbb{A}^{-1}$ corresponds to raising its second index. Moreover, since~$\accentset{\circ}{c}_{\alpha\beta}=\accentset{\circ}{b}_{\alpha\gamma}\accentset{\circ}{b}^{\gamma}_{\,\,\beta}$, we can write, in matrix notation, $\mathbb{C}=\mathbb{B}\mathbb{A}^{-1}\mathbb{B}$. With these results, we can rewrite Eq.~\eqref{eq:Gcontramatrix} in index notation as
\begin{equation}
    G^{\alpha\beta}=\accentset{\circ}{a}^{\alpha\beta}+2\eta^3\accentset{\circ}{b}^{\alpha\beta}+3(\eta^3)^2\accentset{\circ}{c}^{\alpha\beta}+O((\eta^3)^3)\,,
    \label{eq:gcontra}
\end{equation}
which expresses the contravariant components of the 3D metric as a function of the thickness coordinate and surface quantities only.

With the above results at hand, we can finally express the deformation gradient, solely in terms of surface quantities, up to second order in the thickness coordinate. By substituting Eqs.(\ref{eq:undefcov}-\ref{eq:defcov}) and Eq.~\eqref{eq:gcontra} into Eq.~\eqref{eq:F3D}, we obtain
\begin{equation}
\begin{aligned}
    \mathbf{F}&=\mathbf{a}_{\alpha}\otimes\accentset{\circ}{\mathbf{a}}^{\alpha}+\mathbf{n}\otimes\accentset{\circ}{\mathbf{n}}\\
    &+\eta^3(-b^{\alpha}_{\,\,\beta}+\accentset{\circ}{b}^{\alpha}_{\,\,\beta})\mathbf{a}_{\alpha}\otimes\accentset{\circ}{\mathbf{a}}^{\beta}\\
    &+(\eta^3)^2(\accentset{\circ}{c}^{\alpha}_{\,\,\beta} -b_{\eta}^{\,\,\alpha}\accentset{\circ}{b}^{\eta}_{\,\,\beta})\mathbf{a}_{\alpha}\otimes\accentset{\circ}{\mathbf{a}}^{\beta}+O((\eta^3)^3)\,,
    \end{aligned}
    \label{eq:Freduced}
\end{equation}
where we used the Weingarten equations~\citep{docarmo2016}, $\accentset{\circ}{\mathbf{n}},_{\alpha}=-\accentset{\circ}{b}_{\alpha}^{\,\,\beta}\accentset{\circ}{\mathbf{a}}_{\beta}$ and~$\mathbf{n},_{\alpha}=-b_{\alpha}^{\,\,\beta}\mathbf{a}_{\beta}$. Note that this expression of the deformation gradient reduces to that of a plate, when $\accentset{\circ}{b}_{\alpha}^{\,\,\beta}=0$ (\textit{i.e.}, flat mid-surface).

We now seek to further simplify Eq.~\eqref{eq:Freduced}, in the limit of thin shells. To do so, we define the principal curvatures in the undeformed and deformed configuration, respectively, as~$(\accentset{\circ}{k}_1,\accentset{\circ}{k}_2)$ and~$(k_1,k_2)$. In the limit of thin linear elastic shells, the curvatures in both the undeformed and deformed configurations have to be small compared to the average thickness of the shell~$\bar{h}$~\citep{Niordson1985,Pietraszkiewicz1980}, that is $\bar{h}\cdot\max{(\accentset{\circ}{k}_1,\accentset{\circ}{k}_2)}\ll1$ and~$\bar{h}\cdot\max{(k_1,k_2)}\ll1$. Therefore, in the limit of thin linear elastic shells, the linear and quadratic terms of the deformation gradient, Eq.~\eqref{eq:Freduced}, are negligible, with the deformation gradient that can finally be expressed as
\begin{equation}
    \mathbf{F}=\mathbf{a}_{\alpha}\otimes\accentset{\circ}{\mathbf{a}}^{\alpha}+\mathbf{n}\otimes\accentset{\circ}{\mathbf{n}}\,.
    \label{eq:F_thinshells}
\end{equation}
This reduced expression of~$\mathbf{F}$, simplified for thin linear elastic shells, will be instrumental to perform the dimensional reduction of the magneto-elastic energy, as will be shown in Section~\ref{sec:dimensionalreduction}.

\subsection{The reduced elastic energy}

Now that the expression of the deformation gradient for thin shells has been derived, the last task needed before moving to the dimensional reduction procedure is to recall the reduced elastic energy, as proposed by Koiter~\citep{koiter_over_1945,Niordson1985} and used by many authors since then~\citep{budiansky_buckling_1972,paulose_buckling_2013,hutchinson_john_w._buckling_2016}. As commonly adapted in the shell literature \citep{Niordson1985}, we define the following mid-surface stretching and bending strain tensors
\begin{equation}
\label{eq:stretchingstrains}
E_{\alpha\beta}= \frac{1}{2}(a_{\alpha\beta}-\accentset{\circ}{a}_{\alpha\beta})\,
\end{equation}

\begin{equation}
\label{eq:bendingstrains}
K_{\alpha\beta}= b_{\alpha\beta}-\accentset{\circ}{b}_{\alpha\beta}\,,
\end{equation}
the elastic energy of a linearly elastic shell can then be expressed as~\citep{Niordson1985}
\begin{equation}
\begin{aligned}
\mathcal{U}_\textup{e}&=\int_\mathcal{S} \frac{Eh}{2(1-\nu^2)}[(1-\nu)E_{\alpha}^{\beta} E_{\beta}^{\alpha}+\nu E_{\alpha}^{\alpha} E_{\beta}^{\beta}]\,\dd\accentset{\circ}{\omega}\\&+\int_\mathcal{S} \frac{Eh^3}{24(1-\nu^2)}[(1-\nu)K_{\alpha}^{\beta} K_{\beta}^{\alpha}+\nu K_{\alpha}^{\alpha} K_{\beta}^{\beta}]\,\dd\accentset{\circ}{\omega}\,,
\end{aligned}
\label{eqn:energy}
\end{equation}
where~$\dd\accentset{\circ}{\omega} = \sqrt{|\mathrm{det}(\accentset{\circ}{a}_{\alpha\beta})|} \,\dd\eta^1\dd\eta^2=\accentset{\circ}{a} \,\dd\eta^1\dd\eta^2$ is the differential area, $E$ is the Young's modulus of the material and~$\nu$ its Poisson ratio. In the test cases that we will focus on in this work, the material properties and the thickness will be assumed to be homogeneous. The elastic energy in Eq.~(\ref{eqn:energy}) can also be written in a more compact direct notation, by introducing the trace operator in the undeformed metric ``$\tr$'' such that for example~$\tr\mathbf{a}=\accentset{\circ}{a}^{\alpha\beta}a_{\alpha\beta}$, so that 
\begin{equation}\label{energy1D}
\begin{aligned}
\mathcal{U}_\textup{e}&=\frac{Eh}{8(1-\nu^2)}\int_\mathcal{S} [(1-\nu)\tr(\vett{a}-\accentset{\circ}{\vett{a}})^2+\nu\tr^2(\vett{a}-\accentset{\circ}{\vett{a}})] \,\dd\accentset{\circ}{\omega}\\&+\frac{D}{2}\int_\mathcal{S} [(1-\nu)\tr(\vett{b}-\accentset{\circ}{\vett{b}})^2+\nu\tr^2(\vett{b}-\accentset{\circ}{\vett{b}})] \,\dd\accentset{\circ}{\omega}\,,
\end{aligned}
\end{equation}
\noindent{
\begin{sloppypar}
where we have also defined the bending stiffness of the shell as ${D=1/12Eh^3/(1-\nu^2)}$.
\end{sloppypar}}

In this Section, we have described the geometry of the shell, writing the deformation gradient for thin linear elastic shells, and recalling some basic concepts in shell mechanics that will be useful for the remaining of the paper, where we will introduce the magnetic Helmholtz free energy and study the coupling between mechanics and magnetism. We are now ready to move on to the next Section, where we will obtain the reduced energy for magneto-active shells.

\section{Reduced magneto-elastic energy via dimensional reduction}
\label{sec:dimensionalreduction}

The goal of this Section is to reduce the dimension of the magnetic energy to obtain a reduced order model for non-axisymmetric thin elastic magnetic shells. This model will provide a much faster alternative to multiphysics 3D simulations, in addition to being amenable to physical interpretation, as we will show in Section~\ref{sec:interpretation}.

The total energy~$\mathcal{U}$ of the shell is the sum of its elastic energy~$\mathcal{U}_\textup{e}$, the potential energy of external loads such as pressure, and the magnetic energy~$\mathcal{U}_\textup{m}$, which we want to reduce from 3D to 2D. The magnetic energy $\mathcal{U}_\textup{m}$ can be expressed as the Helmholtz free energy for ideal hard-magnetic soft materials~\citep{kim_printing_2018,Zhao2019}, and is inherently 3D, meaning that it is an energy per unit volume. The decomposition of the total energy into elastic and non elastic terms is also found in other contexts, as in the case of the swelling of hydrogels with the Flory-Rehner energy~\citep{Flory1943a,Flory1943b}. This decomposition ensures that the elastic part can be linked to any of the most common constitutive models, including the Kirchhoff Saint-Venant strain energy used in the Koiter shell model or the neo-Hookean energy commonly used for elastomeric materials.

For the magnetic Helmholtz free energy, we will consider the model proposed by~\cite{Zhao2019}, which is valid for ideal hard-magnetic soft materials. The basic underlying assumption, following the physical observations in~\citep{Bertotti_magnetism_1998}, is that the magnetic flux density~$\mathbf{B}$ of the hard-magnetic soft material in the reference configuration is linearly related to the applied magnetic field~$\mathbf{H}$. This assumption is generally true for hard-magnetic soft materials, where the field strength required for actuation is much lower than the coercivity~\citep{Bertotti_magnetism_1998}.

The magnetic energy is then simplified as the work required to align the residual magnetic moment of the material along the external magnetic field for 3D scale-free materials, written as~\citep{Bertotti_magnetism_1998,kim_printing_2018,Zhao2019}%
\begin{equation}
    \mathcal{U}_\textup{m}=-\frac{1}{\mu_0}\int_\mathcal{B} \mathbf{F}\mathbf{B}^\textup{r}\cdot\mathbf{B}^\textup{a}\ \textup{d}V\,,
    \label{eq:mage}
\end{equation}
where~$\mu_0$ is the vacuum permeability, $\mathbf{B}^\textup{r}$ is the vector of residual magnetic flux density, $\mathbf{B}^\textup{a}$ is the vector of externally applied magnetic flux density, and~$V$ is the reference volume of the 3D body $\mathcal{B}$. 

In preparation for the dimensional reduction procedure, we now need to represent the two magnetic fields, namely the residual magnetic flux density and the applied magnetic flux density, via their Cartesian components, noting that they can be functions of the spatial coordinates: $\mathbf{B}^\textup{r}={{B}^\textup{r}}^i\mathbf{e}_i$ and~$\mathbf{B}^\textup{a}={{B}^\textup{a}}^i\mathbf{e}_i$. We then define the normalized components as
\begin{equation}
\begin{aligned}
{\hat{B}^\textup{r}}^i&=\frac{{B^\textup{r}}^i}{|\mathbf{B}^\textup{r}|}\,,\\
{\hat{B}^\textup{a}}^i&=\frac{{B^\textup{a}}^i}{|\mathbf{B}^\textup{a}|}\,.
\end{aligned}
\end{equation}
Then, substituting the expressions for the deformation gradient (Eq.~\eqref{eq:F_thinshells}) and for the volume measure (Eq.~\eqref{eq:jacobian}) in the 3D magnetic energy (Eq.~\eqref{eq:mage}), and integrating along the thickness, yields
\begin{equation}
\mathcal{U}_\textup{m}=hE\lambda_\textup{m}\int_{\mathcal{S}}  B^{ij}_0[(\mathbf{a}_{\alpha}\cdot\mathbf{e}_j)(\accentset{\circ}{\mathbf{a}}^{\alpha}\cdot\mathbf{e}_i)+(\mathbf{n}\cdot\mathbf{e}_j)(\accentset{\circ}{\mathbf{n}}\cdot\mathbf{e}_i)]\,\dd\accentset{\circ}{\omega}\,,
\label{eq:magenergy}
\end{equation}
where, given the thin shell assumption, we neglected~$h\accentset{\circ}{H}$ and~$h^2\accentset{\circ}{K}$ compared to unity. Indeed, the mean curvature is of the order of the principal curvatures of the shell, while the Gaussian curvature is equal to their product. Recall that, in Section \ref{sec:preliminaries}, we highlighted that the assumptions on the undeformed shape of the shell are $h\cdot\max{(\accentset{\circ}{k}_1,\accentset{\circ}{k}_2)}\ll1$, and, hence, $h\accentset{\circ}{H}\ll1$ and~$h^2\accentset{\circ}{K}\ll1$. Moreover, according to~\cite{Yan_NatureMaterials_2020}, we use the magneto-elastic dimensionless parameter
\begin{equation}
\lambda_\textup{m}=\frac{|\mathbf{B}^\textup{r}||\mathbf{B}^\textup{a}|}{\mu_0E}\,,
\end{equation}
which represents the ratio between an equivalent magnetic pressure and the Young's modulus. If $\lambda_\textup{m}\gg1$, the shell will be \emph{magnetically compliant}, whereas if $\lambda_\textup{m}\ll1$, it will be \emph{magnetically stiff}. Furthermore, we have defined
\begin{equation}
B_0^{ij}=-\frac{1}{h}\int_{-h/2}^{h/2} {\hat{B}^\textup{r}}^i{\hat{B}^\textup{a}}^j\,\textup{d}\eta^3\,,
\end{equation}
which can be seen as the Cartesian components of the~$\mathbf{B}_0=B_0^{ij}\mathbf{e}_i\otimes\mathbf{e}_j$ tensor. The components~$B_0^{ij}$ are the average of the product~${\bar{B}^\textup{r}}^i{\bar{B}^\textup{a}}^j$ along the thickness of the shell, and quantify the interaction between the residual and the externally applied magnetic flux densities.

The reduced energy in Eq.~\eqref{eq:magenergy} is now suitable for the modeling of thin shells made of hard MREs subject to external magnetic fields, which can be either homogeneous or linear (constant gradient fields). 

In the next Section, we will investigate the reduced energy to gain additional insight and obtain a physical interpretation of how the external magnetic field acts to deform the shell.

\section{Reduced magneto-elastic energy as load potentials}
\label{sec:interpretation}

The reduced magnetic energy in Eq.~\eqref{eq:magenergy} is nonlinear and, therefore, opaque to an intuitive understanding on how the magnetic field interacts with the mechanics of the shell. Therefore, we seek to provide a physical interpretation of the newly reduced magnetic energy, in the case where both the residual and the applied magnetic fields are homogeneous. Previously, we presented a similar approach, albeit limited to axisymmetric deformations, in~\citep{Yan_NatureMaterials_2020}. The goal of the current Section is to expand the reduced magnetic energy in Eq.~\eqref{eq:magenergy} as a function of the displacement field and to interpret it as the potential energy of \emph{magnetic loads}.

We start by expanding the reduced magnetic energy up to second order in the displacement field~$\mathbf{u}$. The term that we need to expand is~$[(\mathbf{a}_{\alpha}\cdot\mathbf{e}_j)(\accentset{\circ}{\mathbf{a}}^{\alpha}\cdot\mathbf{e}_i)+(\mathbf{n}\cdot\mathbf{e}_j)(\accentset{\circ}{\mathbf{n}}\cdot\mathbf{e}_i)]$. We write~$\mathbf{a}_{\alpha}=\accentset{\circ}{\mathbf{a}}_{\alpha}+\delta\mathbf{a}_{\alpha}$ and~$\mathbf{n}=\accentset{\circ}{\mathbf{n}}+\delta\mathbf{n}$, where $\delta\mathbf{a}_{\alpha}$ and $\delta\mathbf{n}$ are the (finite) changes in the base vectors corresponding to a (finite) displacement $\mathbf{u}$, and notice that
\begin{equation}
(\mathbf{a}_{\alpha}\cdot\mathbf{e}_j)(\accentset{\circ}{\mathbf{a}}^{\alpha}\cdot\mathbf{e}_i)+(\mathbf{n}\cdot\mathbf{e}_j)(\accentset{\circ}{\mathbf{n}}\cdot\mathbf{e}_i)=\delta_{ij}+(\delta\mathbf{a}_{\alpha}\cdot\mathbf{e}_j)(\accentset{\circ}{\mathbf{a}}^{\alpha}\cdot\mathbf{e}_i)+(\delta\mathbf{n}\cdot\mathbf{e}_j)(\accentset{\circ}{\mathbf{n}}\cdot\mathbf{e}_i)\,,
\label{eq:enlinear}
\end{equation}
where~$\delta_{ij}$ is the Kronecker's delta.

Then, since~$\mathbf{r}=\accentset{\circ}{\mathbf{r}}+\mathbf{u}$, we can derive the expression for $\delta\mathbf{a}_{\alpha}$ as follows
\begin{equation}
\begin{aligned}
    \delta\mathbf{a}_{\alpha}&=(\mathbf{r}-\accentset{\circ}{\mathbf{r}}),_{\alpha}=(u^{\gamma}\accentset{\circ}{\mathbf{a}}_{\gamma}),_{\alpha}+(w\accentset{\circ}{\mathbf{n}}),_{\alpha}\\
    &=u^{\gamma},_{\alpha}\accentset{\circ}{\mathbf{a}}_{\gamma}+u^{\gamma}\accentset{\circ}{\mathbf{a}}_{\gamma},_{\alpha}+w,_{\alpha}\accentset{\circ}{\mathbf{n}}-w\accentset{\circ}{b}_{\alpha}^{\,\,\beta}\accentset{\circ}{\mathbf{a}}_{\beta}\\
    &=(u^{\eta},_{\alpha}+\Gamma^{\eta}_{\,\,\gamma\alpha}u^{\gamma})\accentset{\circ}{\mathbf{a}}_{\eta}+(\accentset{\circ}{b}_{\gamma\alpha}u^{\gamma}+w,_{\alpha})\accentset{\circ}{\mathbf{n}}-w\accentset{\circ}{b}_{\alpha}^{\,\,\beta}\accentset{\circ}{\mathbf{a}}_{\beta}\\
    &=(\nabla_{\alpha}u^{\eta}-w\accentset{\circ}{b}_{\alpha}^{\,\,\eta})\accentset{\circ}{\mathbf{a}}_{\eta}+(\accentset{\circ}{b}_{\gamma\alpha}u^{\gamma}+w,_{\alpha})\accentset{\circ}{\mathbf{n}}\,,
    \end{aligned}
    \label{eq:deltaa}
\end{equation}
where we used the Weingarten equation~$\accentset{\circ}{\mathbf{n}},_{\alpha}=-\accentset{\circ}{b}_{\alpha}^{\,\,\beta}\accentset{\circ}{\mathbf{a}}_{\beta}$, $\Gamma^{\eta}_{\,\,\gamma\alpha}=\accentset{\circ}{\mathbf{a}}_{\gamma},_{\alpha}\cdot\accentset{\circ}{\mathbf{a}}^{\eta}$ is the Christoffel symbol of the second kind, and~$\nabla_{\alpha}$ denotes the covariant derivative along~$\accentset{\circ}{\mathbf{a}}_{\alpha}$~\citep{docarmo2016}. We note that this expression is geometrically exact, given that the covariant base vectors are linear functions of the displacement field and its derivatives.

On the contrary, the normal vector is by definition a nonlinear function of the displacement field. The expansion of $\delta\mathbf{n}$ up to second order reads~\citep{Deserno2004}
\begin{equation}
    \delta\mathbf{n}=-\frac{1}{2}q_{\alpha}q^{\alpha}\accentset{\circ}{\mathbf{n}}-(q^{\alpha}-q^{\gamma}U^{\,\,\alpha}_{\gamma})\accentset{\circ}{\mathbf{a}}_{\alpha}+O(|\mathbf{u}|^3)\,,
    \label{eq:deltan}
\end{equation}
where~$q_{\alpha}=w,_{\alpha}+\accentset{\circ}{b}_{\gamma\alpha}u^{\gamma}$ are the covariant components of the rotation vector~\citep{Niordson1985}, and~$U^{\,\,\alpha}_{\gamma}=\nabla_{\gamma}u^{\alpha}-w\accentset{\circ}{b}_{\gamma}^{\,\,\alpha}$ are the mixed components of the surface gradient of the displacement field, that is~$\nabla_{\!\mathcal{S}}\mathbf{u}$.

Finally, substituting Eqs.~\eqref{eq:deltaa}, \eqref{eq:deltan} and~\eqref{eq:enlinear} into \eqref{eq:magenergy}, we get
\begin{equation}
\label{eq:expansion}
\begin{aligned}
    \mathcal{U}_\textup{m}^2&=hE\lambda_\textup{m}\int_\mathcal{S}B_0^{ij}\delta_{ij}\,\dd\accentset{\circ}{\omega}\\
    +h&E\lambda_\textup{m}\int_\mathcal{S}B_0^{ij}U_{\alpha}^{\,\,\eta}(\accentset{\circ}{\mathbf{a}}_{\eta}\cdot\mathbf{e}_j)(\accentset{\circ}{\mathbf{a}}^{\alpha}\cdot\mathbf{e}_i)\,\dd\accentset{\circ}{\omega}
    -hE\lambda_\textup{m}\int_\mathcal{S}B_0^{ij}\frac{1}{2}q_{\alpha}q^{\alpha}(\accentset{\circ}{\mathbf{n}}\cdot\mathbf{e}_j)(\accentset{\circ}{\mathbf{n}}\cdot\mathbf{e}_i)\,\dd\accentset{\circ}{\omega}\\+h&E\lambda_\textup{m}\int_\mathcal{S}B_0^{ij}q^{\gamma}U^{\alpha}_{\,\,\gamma}(\accentset{\circ}{\mathbf{a}}_{\alpha}\cdot\mathbf{e}_j)(\accentset{\circ}{\mathbf{n}}\cdot\mathbf{e}_i)\,\dd\accentset{\circ}{\omega}\,,
\end{aligned}
\end{equation}
where $\mathcal{U}_\textup{m}^2$ denotes the reduced magnetic energy up to second order in the displacement field. This expansion represents the first step towards a mechanical interpretation of the reduced magnetic energy, where the aim is to understand how the magnetic field loads the shell, in terms of equivalent forces and torques. In the next two subsections, we will analyze each term (apart from the constant term in the first line of Eq.~\eqref{eq:expansion}), and aim at manipulating them to understand their physical meaning. For convenience, we rewrite Eq.~\eqref{eq:expansion} as
\begin{equation}
    \mathcal{U}_\textup{m}^2= \mathcal{U}_\textup{m}^1+\mathcal{U}_\textup{m}^q+\mathcal{U}_\textup{m}^{\tau}\,,
    \label{eq:ensplit}
\end{equation}
where we define
\begin{equation}
\label{eq:terms}
\begin{aligned}
    \mathcal{U}_\textup{m}^1&=hE\lambda_\textup{m}\int_\mathcal{S}B_0^{ij}U_{\alpha}^{\,\,\eta}(\accentset{\circ}{\mathbf{a}}_{\eta}\cdot\mathbf{e}_j)(\accentset{\circ}{\mathbf{a}}^{\alpha}\cdot\mathbf{e}_i)\,\dd\accentset{\circ}{\omega}\,,\\
    \mathcal{U}_\textup{m}^q&=-hE\lambda_\textup{m}\int_\mathcal{S}B_0^{ij}\frac{1}{2}q_{\alpha}q^{\alpha}(\accentset{\circ}{\mathbf{n}}\cdot\mathbf{e}_j)(\accentset{\circ}{\mathbf{n}}\cdot\mathbf{e}_i)\,\dd\accentset{\circ}{\omega}\,,\\
     \mathcal{U}_\textup{m}^{\tau}&=hE\lambda_\textup{m}\int_\mathcal{S}B_0^{ij}q^{\gamma}U^{\alpha}_{\,\,\gamma}(\accentset{\circ}{\mathbf{a}}_{\alpha}\cdot\mathbf{e}_j)(\accentset{\circ}{\mathbf{n}}\cdot\mathbf{e}_i)\,\dd\accentset{\circ}{\omega}\,.
     \end{aligned}
\end{equation}
We will first tackle the linear term $\mathcal{U}_\textup{m}^1$ and then move on to the two second order terms, namely $\mathcal{U}_\textup{m}^q$ and $\mathcal{U}_\textup{m}^{\tau}$.

\subsection{Linear term of the reduced magnetic energy, $\mathcal{U}_\textup{m}^1$}

We expand the linear term, $\mathcal{U}_\textup{m}^1$, as
\begin{equation}
\begin{aligned}
\mathcal{U}_\textup{m}^1&=hE\lambda_\textup{m}\int_\mathcal{S}\Bigl[\nabla_{\alpha}u^{\eta}B_0^{ij}(\accentset{\circ}{\mathbf{a}}_{\eta}\cdot\mathbf{e}_j)(\accentset{\circ}{\mathbf{a}}^{\alpha}\cdot\mathbf{e}_i)-w\accentset{\circ}{b}_{\alpha}^{\,\,\eta}B_0^{ij}(\accentset{\circ}{\mathbf{a}}_{\eta}\cdot\mathbf{e}_j)(\accentset{\circ}{\mathbf{a}}^{\alpha}\cdot\mathbf{e}_i)\Bigr]\,\dd\accentset{\circ}{\omega}\,,\\
&=hE\lambda_\textup{m}\int_\mathcal{S}\Bigl[\nabla_{\alpha}u^{\eta}B_0^{ij}(\accentset{\circ}{\mathbf{a}}_{\eta}\cdot\mathbf{e}_j)(\accentset{\circ}{\mathbf{a}}^{\alpha}\cdot\mathbf{e}_i)-\frac{1}{h}p_\textup{mag}w\Bigr]\,\dd\accentset{\circ}{\omega}\,,
\end{aligned}
\label{eq:linearterms}
\end{equation}
where we defined an equivalent (dimensionless) magnetic pressure
\begin{equation}
    p_\textup{mag}=h\accentset{\circ}{b}_{\alpha}^{\,\,\eta}B_0^{ij}(\accentset{\circ}{\mathbf{a}}_{\eta}\cdot\mathbf{e}_j)(\accentset{\circ}{\mathbf{a}}^{\alpha}\cdot\mathbf{e}_i)=h\accentset{\circ}{\mathbf{b}}\colon\!\mathbf{B}_0\,,
\end{equation}
which we can also express as the double inner product between the curvature tensor~$\accentset{\circ}{\mathbf{b}}=\accentset{\circ}{b}_{\alpha\beta}\accentset{\circ}{\mathbf{a}}^{\alpha}\otimes\accentset{\circ}{\mathbf{a}}^{\beta}$ and the (dimensionless) magnetic tensor~$\mathbf{B}_0=B_0^{ij}\mathbf{e}_i\otimes\mathbf{e}_j$. To interpret the first term of the r.h.s. in Eq.~\eqref{eq:linearterms}, we need to expand the covariant derivative and integrate by parts, as follows
\begin{equation}
\begin{aligned}
    &\int_\mathcal{S}\nabla_{\alpha}u^{\eta}B_0^{ij}(\accentset{\circ}{\mathbf{a}}_{\eta}\cdot\mathbf{e}_j)(\accentset{\circ}{\mathbf{a}}^{\alpha}\cdot\mathbf{e}_i)\accentset{\circ}{a} \,\dd\eta^1\dd\eta^2\\
    &=\int_\mathcal{S}[u^{\eta},_{\alpha}B_0^{ij}(\accentset{\circ}{\mathbf{a}}_{\eta}\cdot\mathbf{e}_j)(\accentset{\circ}{\mathbf{a}}^{\alpha}\cdot\mathbf{e}_i)+\Gamma_{\gamma\alpha}^{\eta}B_0^{ij}(\accentset{\circ}{\mathbf{a}}_{\eta}\cdot\mathbf{e}_j)(\accentset{\circ}{\mathbf{a}}^{\alpha}\cdot\mathbf{e}_i)u^{\gamma}]\accentset{\circ}{a}\,\dd\eta^1\dd\eta^2\\
    &=\int_\mathcal{S}(u^{\eta}B_0^{ij}(\accentset{\circ}{\mathbf{a}}_{\eta}\cdot\mathbf{e}_j)(\accentset{\circ}{\mathbf{a}}^{\alpha}\cdot\mathbf{e}_i)\accentset{\circ}{a}),_{\alpha}\,\dd\eta^1\dd\eta^2+\frac{1}{h}\int_\mathcal{S} f_{\gamma}u^{\gamma}\accentset{\circ}{a}\,\dd\eta^1\dd\eta^2\,.
 \end{aligned}
 \label{eq:force}
\end{equation}   
Here, we have defined an equivalent (dimensionless) membrane distributed force $\mathbf{f}_\textup{mag}=~f_{\gamma}\accentset{\circ}{\mathbf{a}}^{\gamma}$ as
 \begin{equation}
 f_{\gamma}=h\Gamma_{\gamma\alpha}^{\eta}B_0^{ij}(\accentset{\circ}{\mathbf{a}}_{\eta}\cdot\mathbf{e}_j)(\accentset{\circ}{\mathbf{a}}^{\alpha}\cdot\mathbf{e}_i)-h\frac{(B_0^{ij}(\accentset{\circ}{\mathbf{a}}_{\gamma}\cdot\mathbf{e}_j)(\accentset{\circ}{\mathbf{a}}^{\alpha}\cdot\mathbf{e}_i)\accentset{\circ}{a}),_{\alpha}}{\accentset{\circ}{a}}\,,
\end{equation}
where the last term is in the form of the covariant divergence of a vector field. When the shells we consider have a fixed boundary or are boundary-free, the divergence term in Eq.~\eqref{eq:force} integrates to zero and we can conclude that
\begin{equation}\label{eq:um1}
\mathcal{U}_\textup{m}^1=E\lambda_\textup{m}(f_{\gamma}u^{\gamma}-p_\textup{mag}w)=E\lambda_\textup{m}(\mathbf{f}_\textup{mag}\cdot\check{\mathbf{u}}-p_\textup{mag}w)\,.
\end{equation}
This equation suggests that, at the first order in the displacement field, the reduced magneto-elastic energy can be interpreted as the energy potential of distributed magnetic in-plane forces $\mathbf{f}_\textup{mag}$ and pressure $p_\textup{mag}$. This interpretation of $\mathcal{U}_\textup{m}^1$ generalizes the one we derived for axisymmetric deformations presented in~\citep{Yan_NatureMaterials_2020}, which also resulted in the definition of magnetic in-plane forces and pressure.

\subsection{Nonlinear terms of the reduced magnetic energy, $\mathcal{U}_\textup{m}^q$ and $\mathcal{U}_\textup{m}^{\tau}$}

We have just provided a mechanical interpretation to the linear term in the reduced magneto-elastic energy, namely $\mathcal{U}_\textup{m}^1$, but we still need to do the same for the second-order terms $\mathcal{U}_\textup{m}^q$ and $\mathcal{U}_\textup{m}^{\tau}$.

We start by looking at $\mathcal{U}_\textup{m}^q$ which, as shown by Eq.~\eqref{eq:terms}, is quadratic in the linear  rotation vector $\mathbf{q}$. Indeed, $\mathcal{U}_\textup{m}^q$ corresponds to the energy potential of a distribution of linear torques throughout the shell, analogous to a distribution of effective rotational springs of constant stiffness~$k$, proposed by~\cite{Yan_NatureMaterials_2020}. This stiffness $k$, in our more general case, can be defined as
\begin{equation}\label{eq:qterm}
k=B_0^{ij}(\accentset{\circ}{\mathbf{n}}\cdot\mathbf{e}_j)(\accentset{\circ}{\mathbf{n}}\cdot\mathbf{e}_i)=(\accentset{\circ}{\mathbf{n}}\otimes\accentset{\circ}{\mathbf{n}})\colon\!\mathbf{B}_0\,,
\end{equation}
which, depending on its sign, can drastically change the nature of the effective rotational springs. As an illustrative example, let us consider a case where $\mathbf{B}^\textup{r}={{B}^\textup{r}}\mathbf{e}_3$ and~$\mathbf{B}^\textup{a}={{B}^\textup{a}}\mathbf{e}_3$, considered also in \citep{Yan_NatureMaterials_2020}. This specific case implies that $k=-\hat{B}^\textup{r}\hat{B}^\textup{a}(\accentset{\circ}{\mathbf{n}}\cdot\mathbf{e}_3)^2$ or, stated differently, that the sign of the stiffness of the rotational springs is opposite to that of $\mathbf{B}^\textup{r}\cdot\mathbf{B}^\textup{a}$. This simple example conveys the importance of such a physical interpretation of the reduced magneto-elastic energy, since a quick check of the sign of $\mathbf{B}^\textup{r}\cdot\mathbf{B}^\textup{a}$ provides a first qualitative assessment of the case or problem at hand.

The last second order term in Eq.~\eqref{eq:ensplit}, namely $\mathcal{U}_\textup{m}^{\tau}$, can be interpreted as the energy potential of distributed torques, which depend on the surface gradient of the displacement field. We define this dimensionless distributed torque as $\boldsymbol{\tau}(\mathbf{u})~=\tau_{\gamma}\accentset{\circ}{\mathbf{a}}^{\gamma}$, where
\begin{equation}\label{eq:tauterm}
    \tau_{\gamma}=B_0^{ij}(\accentset{\circ}{\mathbf{a}}_{\alpha}\cdot\mathbf{e}_j)(\accentset{\circ}{\mathbf{n}}\cdot\mathbf{e}_i)U^{\alpha}_{\,\,\gamma}\,
\end{equation}
is a linear function of the displacement field, since $U^{\alpha}_{\,\,\gamma}$ is linear with the displacement field.

Using the above results on the mechanical interpretation, namely Eqs. \eqref{eq:um1}, \eqref{eq:qterm} and \eqref{eq:tauterm}, the reduced magnetic energy up to second order in the displacement field reads
\begin{equation}
\mathcal{U}_\textup{m}^2=E\lambda_\textup{m}\int(\mathbf{f}_\textup{mag}\cdot\check{\mathbf{u}}-p_\textup{mag}w)\,\dd\accentset{\circ}{\omega}-\frac{1}{2}hE\lambda_\textup{m}\int k\mathbf{q}\cdot\mathbf{q}\,\dd\accentset{\circ}{\omega}+hE\lambda_\textup{m}\int\boldsymbol{\tau}(\mathbf{u})\cdot\mathbf{q}\,\dd\accentset{\circ}{\omega}\,,
\label{eq:eneinterpretation}
\end{equation}
for homogeneous magnetic fields and boundary-free (or clamped) shells. Moreover, we recall that the above expression (Eq.~\eqref{eq:eneinterpretation}) of the reduced magnetic energy is geometrically exact up to second order in the displacement field, and that we assumed homogeneous Young's modulus and thickness throughout. In summary, we have shown that the reduced magneto-elastic energy is equivalent to an energy potential of \emph{magnetic loads}, combining in-plane forces, linear pressure, and torques, distributed across the mid-surface of the shell.

In the next Section, we will contrast this second-order energy against the fully nonlinear version, that is~Eq.~\eqref{eq:magenergy}, to understand whether $\mathcal{U}_\textup{m}^2$ represents a good approximation and whether there are terms in the expansion that predominate over others.

\section{Numerical implementation}
\label{sec:numerical}

\subsection{Minimization of the reduced 2D energy}
\label{subsec:numerical}

The reduced magneto-elastic energy in Eq.~\eqref{eq:magenergy} is fully nonlinear and, as with the majority of problems involving thin shells, its minimization (together with the elastic energy) has to be tackled via numerical methods. The mechanical interpretation provided in Section~\ref{sec:interpretation} can serve as a tool to understand the basics of a specific problem and can guide its solution, but quantitative solutions can only be obtained via the numerical minimization of the reduced magneto-elastic energy (Eq.~\eqref{eq:magenergy}). In particular cases, as we will see in Section~\ref{sec:validation}, the numerical minimization of $\mathcal{U}_\textup{m}^2$ might suffice to obtain a satisfactory solution of a problem, resulting in an even faster numerical minimization scheme given that $\mathcal{U}_\textup{m}^2$ is only quadratic in $\mathbf{u}$.

The total energy of the shell $\mathcal{U}$ is the sum of the elastic energy $\mathcal{U}_\textup{e}$, the potential of the live pressure $p\Delta\mathcal{V}$ (with $\Delta\mathcal{V}$ being the change in volume of the shell), and the reduced magnetic energy $\mathcal{U}_\textup{m}$. For our numerical scheme, we nondimensionalize the total energy by $EhR^2/(8(1-\nu^2))$ to obtain

\begin{equation}\label{eq:totalenergy}
\overline{\mathcal{U}}=\overline{\mathcal{U}}_\textup{s}+\frac{1}{3}\left(\frac{h}{R}\right)^2\overline{\mathcal{U}}_\textup{b}+\frac{p}{E}\Delta\overline{\mathcal{V}}+\lambda_\textup{m}\overline{\mathcal{U}}_\textup{m}\,,
\end{equation}
where~$\overline{\mathcal{U}}_\textup{s}$, $\overline{\mathcal{U}}_\textup{b}$, and~$\Delta\overline{\mathcal{V}}$ are the dimensionless stretching energy, bending energy, and change in volume. We can write these terms as~\citep{Pezzulla_JAM_2019}
\begin{equation}
\begin{aligned}
\overline{\mathcal{U}}_\textup{s}&=\frac{1}{R^2}\int [(1-\nu)\tr(\mathbf{a}-\accentset{\circ}{\mathbf{a}})^2+\nu\tr^2(\mathbf{a}-\accentset{\circ}{\mathbf{a}})]\,\dd\accentset{\circ}{\omega}\,,\\
\overline{\mathcal{U}}_\textup{b}&=\int [(1-\nu)\tr(\mathbf{b}-\accentset{\circ}{\mathbf{b}})^2+\nu\tr^2(\mathbf{b}-\accentset{\circ}{\mathbf{b}})]\,\dd\accentset{\circ}{\omega}\,,\\
\Delta\overline{\mathcal{V}}&=\frac{8(1-\nu^2)}{3hR^2}\biggl[\int\mathbf{r}\cdot\mathbf{n}\,\dd\omega-\int\accentset{\circ}{\mathbf{r}}\cdot\accentset{\circ}{\mathbf{n}}\,\dd\accentset{\circ}{\omega}\biggr]\,,\\
\overline{\mathcal{U}}_\textup{m}&=\frac{8(1-\nu^2)}{R^2}\int  B^{ij}_0[(\mathbf{a}_{\alpha}\cdot\mathbf{e}_j)(\accentset{\circ}{\mathbf{a}}^{\alpha}\cdot\mathbf{e}_i)+(\mathbf{n}\cdot\mathbf{e}_j)(\accentset{\circ}{\mathbf{n}}\cdot\mathbf{e}_i)]\,\dd\accentset{\circ}{\omega}\,,
\end{aligned}
\end{equation}
where we added the (dimensionless) reduced magneto-elastic energy $\overline{\mathcal{U}}_\textup{m}$ derived in Section~\ref{sec:dimensionalreduction}. The energy written in Eq.~\eqref{eq:totalenergy} is minimized numerically using the commercial software COMSOL Multiphysics, similarly to how we described in~\citep{Pezzulla_JAM_2019}, even though the procedure presented there was applied to a purely elastic energy, without any magnetic contribution, and to a~$1$D functional since the model was axisymmetric. To discretize the domain of parametrization of the undeformed mid-surface, we have made use of triangular elements using Argyris shape functions~\citep{argyris1968}, a fifth-order polynomial approximation that allows the derivatives of the state variables to be assigned at the boundaries.

\begin{figure}[h]
    \centering
    \includegraphics[scale=0.55]{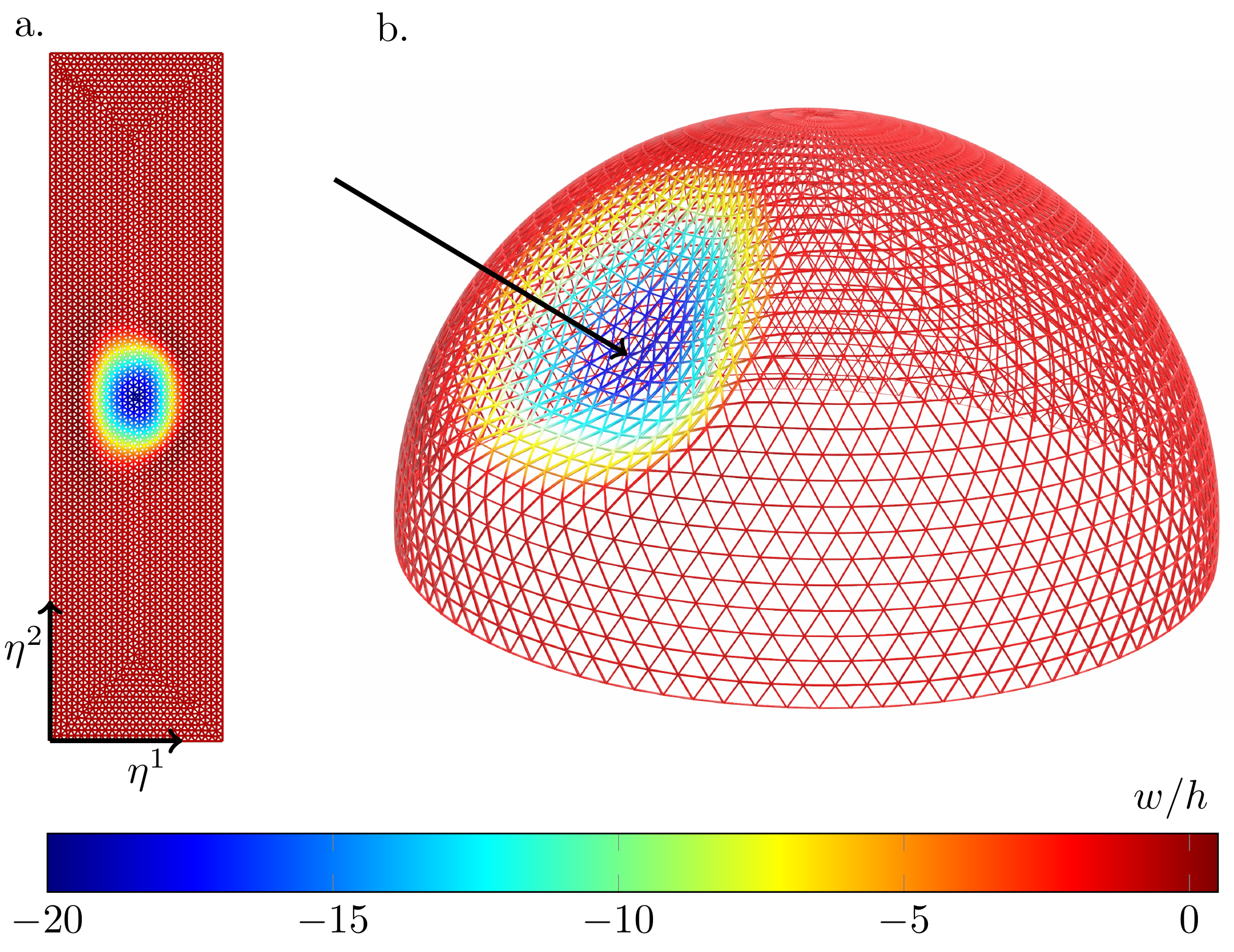}
    \caption{Domain of parametrization for a spherical shell using spherical coordinates $\eta^1\in[0,\pi]$ and $\eta^2\in[0,2\pi)$ (a). The color indicates the dimensionless normal displacement~$w/h$ as a consequence of a poking normal force at a point along the parallel at~$45^\circ$. (b) Deformed mid-surface of the shell in space.}
    \label{fig:mesh}
\end{figure}

In figure~\ref{fig:mesh}(a), we present a representative example of a discretized domain of parametrization $(\eta^1,\eta^2)$ for a hemispherical shell clamped at its equator. In this example, the curvilinear coordinates $({\eta^1,\eta^2})$ are spherical coordinates with~$\eta^1\in[0,\pi/2]$ for the colatitude and~$\eta^2\in[0,2\pi)$ for the longitude. The color map indicates the dimensionless normal displacement, $w/h$, induced by a poking force normal to the surface at a point along the parallel at~$45^\circ$. The corresponding discretized deformed mid-surface in Euclidean space is shown in figure~\ref{fig:mesh}(b).  

In all the examples that we will present in Section~\ref{sec:validation}, we will consider hemispherical shells, clamped along the equator, in some cases with defects precisely engineered on their surface~\citep{lee_geometric_2016}. This spherical geometry corresponds to the special case with $u^{\alpha}=0$, $w=0$, and the clamped boundary translates to $w,_1=0$ at the equator ($\eta^1=\pi/2$)~\citep{Niordson1985}. Moreover, continuity is enforced between the boundary segments at~$\eta^2=0$ and~$\eta^2=2\pi$.


\subsection{FEM simulations of the 3D model}

Our reduced magnetic shell model presented in Secs.~\ref{sec:preliminaries} and~\ref{sec:dimensionalreduction} and its numerical implementation presented above in Sec.~\ref{subsec:numerical} will be validated in Sec.~\ref{sec:validation} using 3D finite element modeling (3D FEM). The 3D FEM was performed in the commercial software package Abaqus/Standard using the user-defined 8-node brick element, proposed by~\cite{Zhao2019} for modeling hard-magnetic deformable solids under a uniform magnetic field. This FEM framework and has been thoroughly validated by experiments~\citep{Zhao2019}. The code was also extended by some of the authors to include the case where the external magnetic field is a constant-gradient field, and successfully applied to study magnetic beams \citep{Yan2020prep}.

The user-defined element that we will use was developed based on the 3D continuum theory of hard-magnetic soft materials~\citep{Zhao2019}. The elastic behavior of the material is described by the neo-Hookean model. The magnetic interaction between the material and externally applied fields is considered by introducing a magnetic part of the Cauchy stress, derived from the 3D magnetic energy potential in Eq.~\eqref{eq:mage}.

In our FEM simulations, a 3D deformable hemispherical shell containing a geometric defect was discretized by the user elements. The material of the shell was assumed to be incompressible, with a bulk modulus 100 times larger than its shear modulus. The shell was fixed at the equator and subjected to combined magnetic and mechanical loading. Contact and distributed mechanical loads (\textit{e.g.}, pressure) were imposed on the shell through a dummy mesh of C3D8R solid elements, which shared the same nodes with the user elements. The dummy material had a negligible elastic modulus ($10^{-20}\,$Pa) compared to the MRE of the shell ($1.8\,$MPa). Geometric nonlinearities were taken into account throughout the simulations. The geometric, material, and loading parameters, provided in Sec.~\ref{sec:validation}, are identical to those used in the shell model. We recall that only problems where the external magnetic field is homogeneous will be considered in Section \ref{sec:validation}, even though both the 3D FEM and the reduced magneto-elastic energy can be used to study problems with constant-gradient magnetic fields. We hope that future work will leverage the intrinsic ability of our theory to consider cases of constant gradient fields.

\section{Validation of the reduced magnetic shell model}
\label{sec:validation}

We proceed by validating the reduced magnetic shell model against the 3D FEM model for magnetorheological elastomers implemented in Abaqus via the study of three different test cases:

\begin{itemize}[noitemsep]
\item[(i)] Non-axisymmetric point indentation under magnetic loading (subsection \ref{ssec:indent}), which will test the model when the magnetic loading is axisymmetric while the mechanical one is not;
\item[(ii)] Pressure buckling under asymmetric magnetic loading where the residual magnetization vector and the external magnetic field are in the same plane (subsection \ref{ssec:asybuckling}), which will test the ability of the model to describe a sub-critical instability under asymmetric magnetic conditions;
\item[(iii)] Pressure buckling under asymmetric magnetic loading where the residual magnetization vector and the external magnetic field are not in the same plane (subsection \ref{ssec:asyopbuckling}), which will further test the model with the study of the elastic instability when the residual and external magnetic fields are non-coplanar.
\end{itemize}

Throughout the validation procedure, and for all test cases, (i), (ii), and (iii), we will also investigate the roles of the reduced energy terms, namely the linear term~$\mathcal{U}_\textup{m}^1$, the second-order torque term (quadratic in the rotation) $\mathcal{U}_\textup{m}^q$, and the full second-order approximation~$\mathcal{U}_\textup{m}^2$. We fix $R=25.2$ mm, $h=0.32$ mm, $E=1.8$ MPa, and $\nu=0.5$, corresponding to the dimensionless numbers $R/h=90$ and $\lambda_\textup{m}=0.00184$. Within all the three test cases, we consider two loading sets identified by the sign of the scalar product between the residual and the applied (homogeneous) magnetic fields, namely $\vett{B}^\textup{r}\cdot\vett{B}^\textup{a}>0$ and $\vett{B}^\textup{r}\cdot\vett{B}^\textup{a}<0$.

\subsection{Point indentation under a magnetic field}
\label{ssec:indent}

The first validation test case considers the asymmetric point indentation of a spherical shell under a uniform and vertical magnetic field, with $\vett{B}^\textup{a}=B^\textup{a}\vett{e}_3$ and $\vett{B}^\textup{r}=B^\textup{r}\vett{e}_3$. The indentation force~$F$ is set normal to the undeformed mid-surface of the shell and located at~$\eta^1=\pi/4$ and~$\eta^2=0$. To include this external force into the reduced shell model, we add its (dimensionless) potential to the total energy in Eq. \eqref{eq:totalenergy}

\begin{equation}
\overline{\mathcal{P}}_\textup{f}=\frac{8(1-\nu^2)}{EhR^2}\int_\mathcal{S} F\delta(\eta^1-\pi/4,\eta^2)w \,\dd\eta^1 \dd\eta^2\,,
\end{equation}
where the force is assumed to be positive if indenting into the shell and $\delta(\eta^1,\eta^2)$ is the Dirac delta function.  

In the 3D FEM simulations, we considered one half of the hemispherical shell (\textit{i.e.}, one quarter of a full spherical shell) with symmetric conditions imposed on the plane spanned by $\vett{e}_1$ and $\vett{e}_3$, in which the indentation was exerted. The indenter was modeled as a rigid sphere of radius $0.2\,$mm and Young's modulus $1.8\,$GPa. The contact between the indenter and the shell was assumed rigid with neither friction nor penetration. We discretized the shell using a swept mesh with 10, 125, 100 seeds, respectively, in the thickness, equatorial, and meridional directions. The mesh in the vicinity of the indentation point was further refined in order to accurately describe the ensuing localized deformation, the characteristic length of which scales as $\sqrt{Rh}$ ~\citep{abbasi_probing_2021}. We have conducted a convergence study to ensure that the results were independent of this discretization. In each simulation, a uniform magnetic field at a given level of flux density (at a fixed value of $\lambda_\textup{m}$) was first applied on the shell, and then, under this fixed field, the shell was indented in a second step.

\begin{figure}[!ht]
    \centering
    \includegraphics[scale=1]{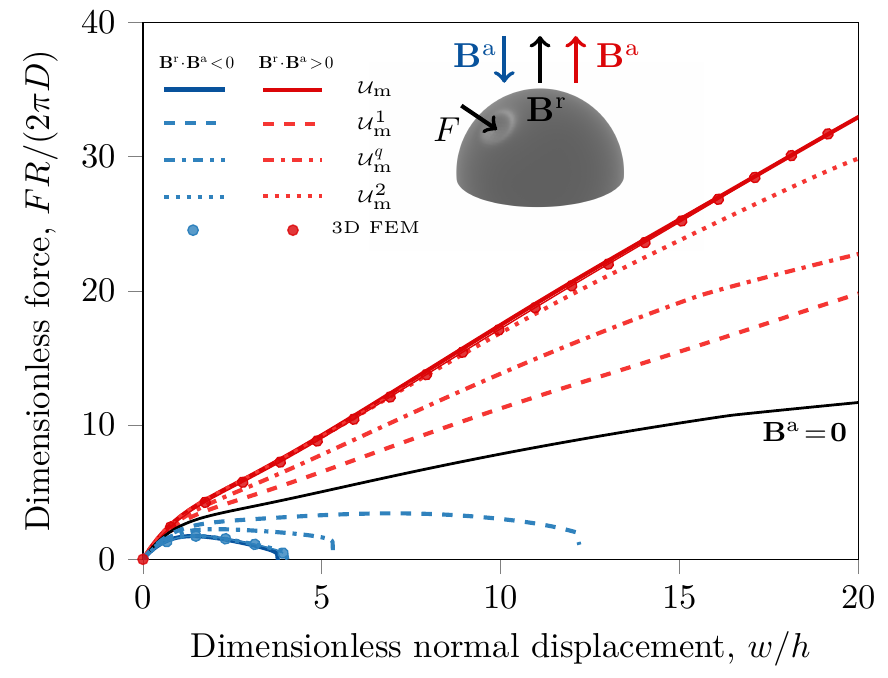}
    \caption{Asymmetric indentation of a spherical shell under a magnetic field in terms of the dimensionless force~$FR/(2\pi D)$ versus the dimensionless normal displacement at the point of indentation $w/h$. The numerical results from the reduced magnetic shell model (solid lines) are in excellent agreement with 3D FEM simulations (symbols). Results in red refer to $\lambda_\textup{m}=0.00184$, $\vett{B}^\textup{r}\cdot\vett{B}^\textup{a}>0$, whereas those in blue refer to $\lambda_\textup{m}=-0.00184$, $\vett{B}^\textup{r}\cdot\vett{B}^\textup{a}<0$. The loading curve for~$\vett{B}^\textup{a}=\vett{0}$ is plotted in black. The results from the second-order reduced energy $\mathcal{U}_\textup{m}^2$, the first-order energy term $\mathcal{U}_\textup{m}^1$, and the torque energy term $\mathcal{U}_\textup{m}^q$ are plotted as dotted lines, dashed lines, and dotted-dashed lines, respectively.}
    \label{fig:indent}
\end{figure}

Figure~\ref{fig:indent} shows the results of the simulations in terms of the dimensionless force~$FR/(2\pi D)$, versus the dimensionless normal displacement at the indentation point, $w/h$. The agreement between our reduced model (solid lines) and the 3D FEM model (symbols) is excellent. Moreover, the results also show how the second-order energy~$\mathcal{U}_\textup{m}^2$ reproduces the results very well, as long as the displacement is not too large (lower than approximately ten times the thickness). By contrast, the first-order energy term $\mathcal{U}_\textup{m}^1$ and second-order torque energy term $\mathcal{U}_\textup{m}^{q}$ fail at replicating the nonlinear behavior, even for small displacements where, consequently, at least a second-order reduced magnetic energy is required. Moreover, from the results in Fig.~\ref{fig:indent}, we notice that the magnetic field modifies the indentation response of the shell similarly to pressure~\citep{Vella2012,Lazarus2012,marthelot_buckling_2017,hutchinson_nonlinear_2017}, since the shell can be either strengthened or weakened depending on whether the pressure is acting to inflate or deflate the shell.

\subsection{Pressure buckling under asymmetric magnetic loading\\ with co-planar $\mathbf{B}^\textup{r}$ and $\mathbf{B}^\textup{a}$}
\label{ssec:asybuckling}

\begin{figure}[!th]
    \centering
    \includegraphics[scale=0.748]{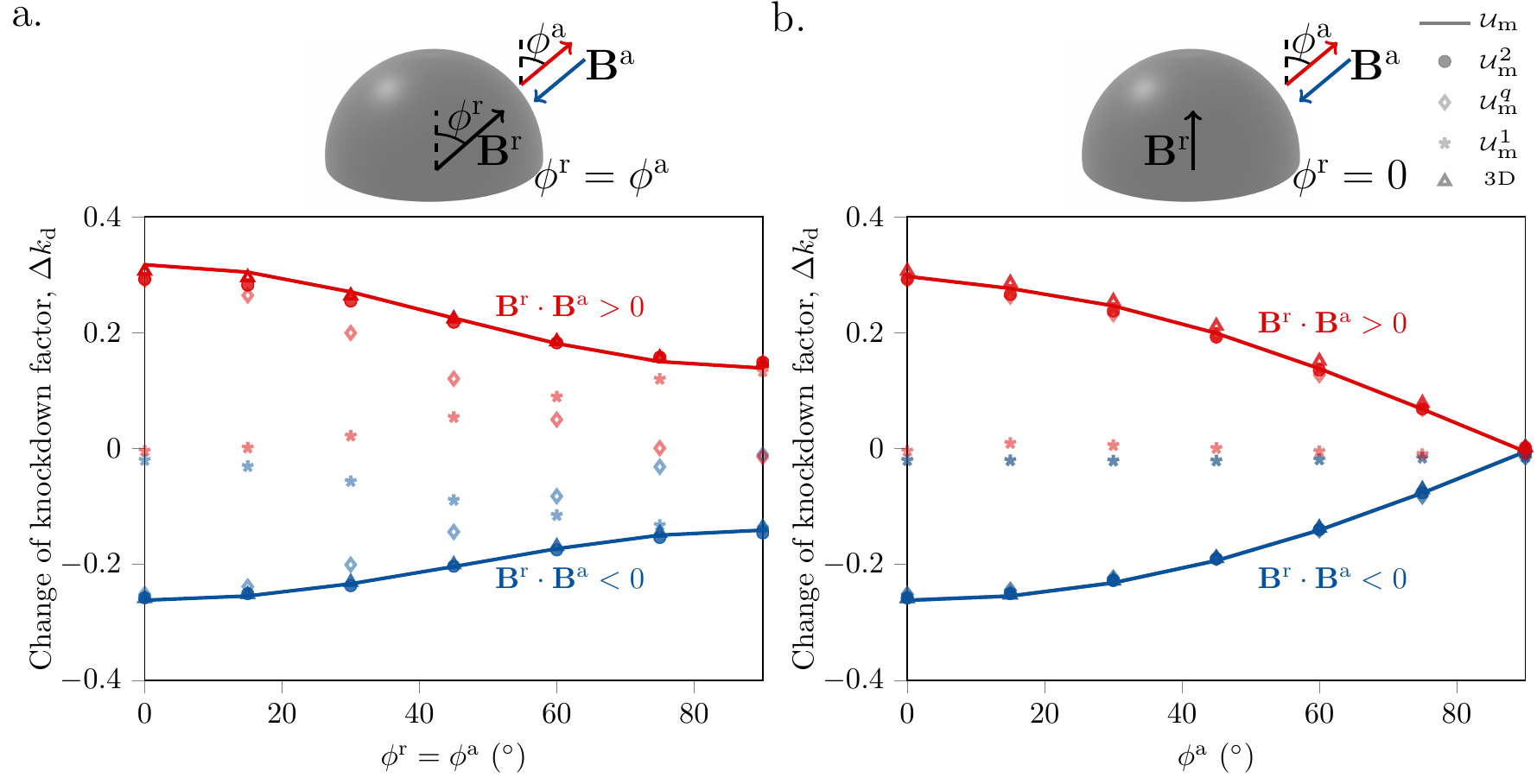}
    \caption{Pressure buckling of a spherical shell under asymmetric magnetic loading for $\lambda_\textup{m}=0.00184$, $\vett{B}^\textup{r}\cdot\vett{B}^\textup{a}>0$ (in red) and $\lambda_\textup{m}=-0.00184$, $\vett{B}^\textup{r}\cdot\vett{B}^\textup{a}<0$ (in blue), with~$\delta/h=1.64$ and~$\bar{\phi}_o=3.2$. (a) Change of the knockdown factor $\Delta k_\textup{d}$ versus the angle of the shell magnetization vector $\phi^\textup{r}$ and of the external applied field $\phi^\textup{a}$, which are consistently aligned. (b) Change of the knockdown factor $\Delta k_\textup{d}$ versus the angle of the external applied field $\phi^\textup{a}$, with $\phi^\textup{r}=0$. Results from the reduced magnetic energy are plotted as solid lines, whereas triangles represent results from the three-dimensional model. Results from the second-order reduced energy $\mathcal{U}_\textup{m}^2$, the first-order energy term $\mathcal{U}_\textup{m}^1$, and the torque energy term $\mathcal{U}_\textup{m}^q$ are plotted as circles, asterisks, and diamonds, respectively.}
    \label{fig:buckling2D}
\end{figure}

The previous test case served to validate our reduced model in the case where the mechanical loading is asymmetric, while the magnetic loading is symmetric. Next, we want to further test our reduced model in the case where the magnetic loading is asymmetric and the shell undergoes an elastic instability, namely pressure-induced buckling. In this test, the residual magnetization vector $\mathbf{B}^\textup{r}$ and the external applied magnetic field $\mathbf{B}^\textup{a}$ are co-planar. If we denote by $\phi^\textup{r}$ and $\phi^\textup{a}$, the angles $\mathbf{B}^\textup{r}$ and $\mathbf{B}^\textup{a}$, respectively, make with $\mathbf{e}_3$, we can distinguish two sub-cases. In the first one, $\mathbf{B}^\textup{r}$ and $\mathbf{B}^\textup{a}$ are parallel, that is $\phi^\textup{r}=\phi^\textup{a}$ (Figure \ref{fig:buckling2D} (a)). In the second one, $\mathbf{B}^\textup{r}$ will stay vertical, that is $\phi^\textup{r}=0$, while $\mathbf{B}^\textup{a}$ will vary via its angle $\phi^\textup{a}$ (Figure \ref{fig:buckling2D} (b)).

Given that shells are highly sensitive to imperfections, their measured critical buckling pressure, even in the absence of any external magnetic fields, can be much lower than classical predictions for a perfect spherical shell~\citep{hutchinson_effect_1971,lee_geometric_2016}. Consequently, it is common to introduce the so-called knockdown factor, $k_\textup{d}$, defined as the ratio between the buckling pressure of the shell and the theoretical buckling pressure of the equivalent perfect shell, $k_\textup{d}=p^\textup{measured}/p_\textup{c}$. Here, we consider a spherical shell, clamped along the equator, with a precisely engineered dimple-like defect at the pole. This defect can be modeled as a deviation from the spherical mid-surface represented, in terms of the normal displacement, as~$w=-\delta(1-(\eta^1/\phi_o)^2)^2$. For the present test case and as a representative example, we set the dimensionless defect amplitude to~$\delta/h=1.64$ and the dimensionless defect width to $\bar{\phi}_o=3.2$, with $\bar{\phi}_o=\phi_o(\sqrt{12(1-\nu^2)}R/h)^{1/2}$. We will present the results in terms of the change of the knockdown factor $\Delta k_\textup{d}$, which is defined as the difference between the knockdown factor obtained when the external magnetic field is on, and that of the shell without any magnetic fields.

This problem has also been studied in~\citep{Yan_NatureMaterials_2020}, where the Abaqus user-defined solid element developed by~\cite{Zhao2019} was employed to predict the critical buckling load of hard-magnetic shells and tested successfully against experiments. Next, we first verify our reduced shell model by comparing with the 3D FEM results presented in~\citep{Yan_NatureMaterials_2020}. Subsequently, using the proposed 2D model, we investigate the relevance of magnetic energy terms at different orders, Eq.~\eqref{eq:expansion}, in shell buckling. 

Figure~\ref{fig:buckling2D}~(a) shows the results of the change in knockdown factor as a function of the angle~$\phi^\textup{r}$, for the case where~$\phi^\textup{r}=\phi^\textup{a}$. The reduced shell model (solid lines) is in excellent agreement with the 3D FEM model (triangles). Moreover, the second-order energy term $\mathcal{U}_\textup{m}^2$ (circles) replicates the results obtained with the full energy almost exactly, being therefore a good approximation of the reduced magnetic energy that can be used for the study of this class of buckling problems. By contrast, neither the first-order energy $\mathcal{U}_\textup{m}^1$ nor the (second-order) torque energy term $\mathcal{U}_\textup{m}^q$ can accurately describe the changes in the knockdown factor. We notice that the first-order energy is adequate to describe the changes in the knockdown factor for $\phi^\textup{r}\simeq\pi/2$, while the torque energy term can be used for $\phi^\textup{r}\simeq 0$. 

In figure~\ref{fig:buckling2D}~(b), we present results for the case when~$\phi^\textup{r}=0$, in terms of the change of the knockdown factor as a function of the angle~$\phi^\textup{a}$. The reduced shell model (solid lines) accurately replicates the results from the 3D FEM model (triangles). Also in this case, the results obtained with the second-order energy term (circles) are in very good agreement with the full energy model. Moreover, we find that the (second-order) torque energy term alone, is able to accurately describe the changes in knockdown factor, while the first-order energy term gives an almost null contribution.

\subsection{Pressure buckling under asymmetric magnetic loading with non co-planar $\mathbf{B}^\textup{r}$ and $\mathbf{B}^\textup{a}$}
\label{ssec:asyopbuckling}

\begin{figure}[!h]
    \centering
    \includegraphics[scale=1]{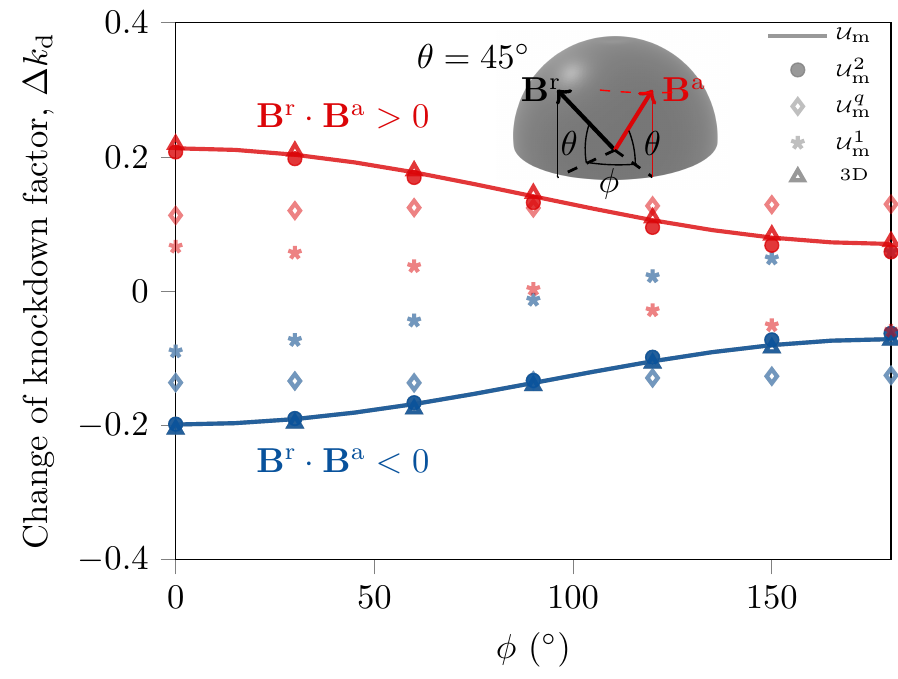}
    \caption{Pressure buckling of a spherical shell under asymmetric magnetic loading, for the more general case when $\vett{B}^\textup{r}$ and $\vett{B}^\textup{a}$ are not in the same plane. $\vett{B}^\textup{r}$ and $\vett{B}^\textup{a}$ are at an angle $\theta=\pi/4$ with respect to the equatorial plane, and separated by an angle $\phi$ along the longitudinal direction. The change of the knockdown factor $\Delta k_\textup{d}$ versus the angle $\phi$ is shown for $\lambda_\textup{m}=0.00184$, $\vett{B}^\textup{r}\cdot\vett{B}^\textup{a}>0$ (in red) and $\lambda_\textup{m}=-0.00184$, $\vett{B}^\textup{r}\cdot\vett{B}^\textup{a}<0$ (in blue). The spherical shells contain a defect at the north pole with dimensionless amplitude~$\delta/h=1.64$ and dimensionless width~$\bar{\phi}_o=3.2$. Results from the second-order reduced energy $\mathcal{U}_\textup{m}^2$ (circles), the first-order energy term $\mathcal{U}_\textup{m}^1$ (asterisks), and the torque energy term $\mathcal{U}_\textup{m}^q$ (diamonds) are also shown.}
    \label{fig:buckling_out}
\end{figure}

Until now, we have validated our reduced shell model in the cases where the magnetic and mechanical loadings are either symmetric or asymmetric, with the residual and external magnetic fields being co-planar. In this last test case, we will evaluate our reduced shell model in the case where the shell undergoes a pressure buckling instability, with $\mathbf{B}^\textup{r}$ and $\mathbf{B}^\textup{a}$ being non co-planar. In particular, as shown in the inset of Fig.~\ref{fig:buckling_out}, both the residual magnetization vector and the external applied field, $\vett{B}^\textup{r}$ and $\vett{B}^\textup{r}$, are at angle $\theta=\pi/4$ with respect to the equatorial plane, and are separated by an angle $\phi$ along the longitudinal direction. Similarly to the previous test cases, the shell contains a defect at the north pole with~$\delta/h=1.64$ and $\bar{\phi}_o=3.2$.

In the 3D FEM of this test case, we modeled the full hemispherical shell with no assumptions of symmetry. The Abaqus simulations were conducted by following the procedure proposed in~\citep{Yan_NatureMaterials_2020} for the pressure buckling of hard-magnetic shells under a uniform field. A swept mesh was created by 8, 400, and 200 seeds in the thickness, equatorial, and meridional directions, respectively, determined through a convergence study. In particular, the seeds in the meridional direction were non-uniform with a bias ratio of 8, such that the mesh was denser near the defect at the north pole, where buckling occurred.


In Fig.~\ref{fig:buckling_out}, we plot the change of the knockdown factor versus the angle $\phi$. We find that the results from the reduced shell model are in excellent agreement with the 3D FEM model. Moreover, the second-order energy term (circles) provides results that agree with the full energy, as in the previous test case discussed in Sec.~\ref{ssec:asybuckling}. Here, the (second-order) torque energy term is not able to describe the changes in the knockdown factor, while the first-order energy term is only successful for $\phi\simeq0$.

\section{Conclusion and discussion}
\label{sec:conclusion}

We have developed a reduced model for magneto-active thin shells with geometrically exact strain measures. Our model is based on a dimensional reduction procedure of the Helmholtz free energy for ideal hard-magnetic soft materials \citep{Zhao2019}. This dimensional reduction relies on the slenderness of the shells ($h/R\ll1$) and on the Kirchhoff-Love assumption on the kinematics, according to which fibers normal to the mid-surface are assumed not to stretch nor shrink and to remain normal to the mid-surface, upon deformation. As a result, the reduced energy for thin shells comprises an elastic energy, split into stretching and bending contributions, the energy potential of the live pressure, and the reduced magnetic energy, which was derived in the present work. This reduced energy is linear in the thickness of the shell, but nonlinear in the displacement field. In its dimensionless form, the reduced magnetic energy introduces a dimensionless parameter $\lambda_\textup{m}$, which we refer to as the magneto-elastic parameter, that can be interpreted as the ratio between the magnetic and the elastic forces in the system.

Since the reduced magnetic energy is highly nonlinear, it is difficult to gain physical insight without additional simplifications. Therefore, we proceeded to expand the energy up to second order in the displacement field $\vett{u}$, for the case where both the residual magnetization field and the external applied field are homogeneous. This expansion yields a first-order energy $\mathcal{U}_\textup{m}^1$, which is related to the work done by equivalent magnetic forces and pressure, and two second-order torque energy terms, with the most predominant term being proportional to the square of the rotation vector, that is $\mathcal{U}_\textup{m}^q$.

We validated our reduced model against a 3D FEM model for MREs and studied three different test cases, involving the indentation and the pressure buckling of spherical shells under magnetic loading. In all the test cases, the results from our reduced model were in excellent agreement with the 3D FEM. Non-trivially, the second-order simplified model was able to replicate the 3D FEM results for the buckling cases and for the indentation case, as long as, in this last case, the normal displacement did not exceed around ten times the thickness. The second-order energy is therefore a good approximation of the reduced magnetic energy for the study of the onsets of the instabilities.

Our model correctly predicts the nonlinear response of thin magnetic shells, when subject to a combination of magnetic, force, and pressure loadings. In all of the three test problems, the reduced model provided solutions that were in excellent agreement with the 3D model, while being an order of magnitude faster than its 3D counterpart. Indeed, we analyzed the computational efficiency of the reduced magnetic shell model with respect to the three-dimensional one, by evaluating the computational times for simulating the indentation problem (figure \ref{fig:indent}) with $20$ points in the load-displacement curves. For both the 3D and 2D models, we measured the computational times when the minimum meshes that ensured convergence were employed. On a Dell Precision 7820 workstation with a 12-core CPU (Intel Xeon Gold 6136) and $192\,$GB RAM, the computational time for the 3D FEM model was $31$ min, whereas the one for the 2D model was $3.5$ min. This tenfold increase in efficiency could be useful for studies where an exploration of a vast parametric space is desired to improve the understanding and design of magnetic shells. Moreover, as we tested the model also in the presence of elastic instabilities, we can conclude that the model is suitable to study those problems where the (subcritical) instabilities, common in thin shells, play a major role. Not only does this reduced model offer a more effective alternative to 3D models, but it also provides a physical insight on the coupling between the magnetic field and the nonlinear mechanics of thin shells. For example, our analysis, applied to the pressure buckling case studied by \cite{Yan_NatureMaterials_2020}, showed how effective rotational springs play a major role in the phenomenon, and revealed the existence of a single governing dimensionless parameter, $\Lambda_\textup{m}=\lambda_\textup{m}R/h$, which summarizes all the geometrical and material properties of the system.

However, there are two main points that are worth of further consideration in future research efforts. The first point regards the self long-range interactions, which have been neglected in both our reduced model and in the 3D FEM model. While these interactions are not significant in the test cases that we investigated for validation purposes, there might be applications were they play a more prominent role. Consider for example a complete spherical shell that is largely deforming under a magnetic field. In this situation, parts of the shell in the northern hemisphere might become close to some parts in the southern hemisphere, making long self-range interactions non-negligible, similarly to magnetic helical rods~\citep{Sano2020prep} in near self-contact. The second point regards a deeper understanding of the simplified first-order and second-order energy terms. Indeed, while our physical interpretation as forces, pressure, and torques simplifies the understanding of the energy, it is not clear why some terms alone are able to describe the deformation of the shell. For example, in the pressure buckling represented in Fig.~\ref{fig:buckling2D}~(b), the second-order torque term $\mathcal{U}_\textup{m}^q$ is sufficient to describe the entire phenomenon, with no clear mechanical explanations.

Our proposed model augments the set of emerging reduced models for slender structures, for planar beams \citep{Yan2020prep,Wang_JMPS2020,Ciambella2018bs,Ciambella2020je} and Kirchhoff-like rods \citep{Sano2020prep}. Moreover, our reduced model extends the shell model presented in \citep{Yan_NatureMaterials_2020}, which was restricted to axisymmetric deformations, now to non-axisymmetric configurations. Such reduced-order models should act as valuable tools for the predictive design of magnetic devices that are becoming increasingly popular in soft robotics \citep{Diller2014fd,Huang2016ee,Pece2017ib,Seffen_SmartMaterStruct2016,Loukaides_IntJSmartNanoMater2014}. Indeed, these models are faster to simulate and more amenable to physical interpretation. We believe that the derivation of a magnetic plate model will be a natural next step towards the further and definitive enrichment of the family of reduced magnetic models. We envision that the different reduced magnetic models could eventually be combined to study complex magnetic structures, comprising rods, plates, and shells elements.

\section*{Acknowledgments}
We are grateful to John W. Hutchinson for fruitful discussions. We thank Alessandro Lucantonio for useful exchanges on the numerical implementation of the model.

\bibliographystyle{elsarticle-harv} 
\bibliography{references.bib}





\end{document}